\documentclass[twocolumn,prl,aps,floatfix]{revtex4} 

\usepackage{graphicx}

\begin{document}


\title{The  magnetic susceptibility of YBa$_{2}$Cu$_{3}$O$_{6+x}$ crystals: unusual Curie behavior and small contributions from charge density waves.}
\author{I. Kokanovi\'{c}$^{1,2}$, J.R. Cooper$^1$}

\affiliation{$^1$Cavendish Laboratory, University of Cambridge, Cambridge CB3 OHE, U.K.\\
 $^2$Department of Physics, Faculty of Science, University of Zagreb, P.O.Box 331,
Zagreb, Croatia.}

\altaffiliation{Electronic address: kivan@phy.hr}



\date{\today}

 \begin{abstract}
We report measurements of the magnetic susceptibility of twinned single crystals of YBa$_{2}$Cu$_{3}$O$_{6+x}$ from just above their superconducting transition temperatures to 300~K with magnetic fields of up to 5~T applied parallel and perpendicular to the CuO$_2$ planes at 7 values of $x$.  Appropriate analysis  allows the relatively small, but still important, Curie terms to be separated from other contributions to the susceptibility. Our data support a picture in which the  Curie terms arise from oxygen disorder in the Cu-O chains.  This agrees with published work  on polycrystalline samples where the sample cooling rate was varied, but here we show that the Curie plots flatten out above 200~K. We identify small effects of  charge density wave (CDW) instabilities in the temperature ($T$) derivative of the in-plane susceptibility $d\chi_{ab}(T)/dT$ and discuss their $x$-dependence. For $x=$0.67 we make a detailed comparison with published high energy X-ray diffraction data using a minimal model involving Fermi arcs, thereby obtaining values for the CDW energy gap and the Helmholtz free energy in a coherence volume. At 80 and 100~K the latter is comparable with, or smaller than  $k_BT$ respectively, highlighting the probable importance of thermal fluctuations. We note that the effect of the Lorentz force on charge carriers in the Fermi arcs could provide a simple mechanism for enhancing the CDWs in high magnetic fields, as suggested by recent experiments.
 \end{abstract}

\maketitle
\section{I. INTRODUCTION}
Nearly ten years ago  the surprising observation of ``slow" quantum oscillations in  under-doped  oxygen-ordered YBa$_2$Cu$_3$O$_{6.5}$ (YBCO) crystals~\cite{Doiron} at high magnetic fields and then in the intrinsically under-doped stoichiometric compound YBa$_2$Cu$_4$O$_8$~\cite{Yelland,Bangura}, opened up new avenues in the study of high temperature  superconductors. More recently, the presence of incommensurate charge density waves (CDWs) at temperatures ($T$) well above the superconducting transition temperature ($T_c$) has been established by a number of experiments, NMR~\cite{Wu2015}, resonant X-ray scattering (RXS)~\cite{Ghiringhelli, Achkar,Blanco} and high energy X-ray diffraction~\cite{Chang} both in YBa$_{2}$Cu$_{3}$O$_{6+x}$, and other underdoped cuprate superconductors~\cite{Comin,Neto,Hucker,Croft}. It has  been found that the wave vector of the CDW is directed along the copper-oxygen bonds and that it decreases slightly with increased hole doping~\cite{Blackburn,Hucker,Blanco,Comin,Croft}.
Angle resolved photoemission~(ARPES) studies~\cite{Comin,Yoshida2012} of  underdoped  bismuth and lanthanum-based cuprates give evidence for Fermi arcs.  In particular Comin et al.~\cite{Comin} suggest that the CDW wave vector gives rise to  small electron pockets~\cite{LeBoeufHall} because it connects the tips of different Fermi arcs. This is appealing because it provides  a direct link between the presence of the pseudogap  and CDW formation. Namely the pseudogap is finite except  in  regions near the diagonals of the Brillouin zone (the nodal regions in the $d$-wave superconducting state). It therefore breaks the  large Fermi surface,  which has been observed directly via quantum oscillation studies of an  overdoped cuprate~\cite{Tl2201}, into disconnected arcs.

 NMR measurements of underdoped YBCO give evidence for
long range static charge order in high fields, without any sign of spin-order for $x>$ 0.5~\cite{Wu2011,Wu2013} while NMR  measurements of quadrupolar frequency broadening  suggest  that the CDW order is also static at low fields~\cite{Wu2015}. Disorder may be important in low fields~\cite{Nie} and on the basis of earlier work on NbSe$_2$, it has been suggested that static CDWs nucleate around defects~\cite{Wu2015}. But somewhat against this, 2$\%$ Zn doping strongly suppresses the CDW intensity in  YBa$_{2}$Cu$_{3}$O$_{6.6}$~\cite{BlancoZn}.
Ultrasonic measurements at high-fields~\cite{LeBoeuf1} have also revealed a transition to a long-range charge ordered state, with further evidence being provided  by two recent structural studies~\cite{Gerber,Chang2015}.  The ionic displacements associated with the CDW  in UD67 YBCO ($T_c$=67~K) have recently been calculated~\cite{Forgan}. They reveal a complicated pattern involving shear displacements perpendicular to the CuO$_2$ layers that breaks the mirror symmetry of the bi-layers.
 For this UD67 crystal  Chang et al.~\cite{Chang}, measured the   intensity of  CDW diffraction peaks
as a function of temperature both at zero and finite magnetic fields.
They observed a gradual increase in intensity  below an onset temperature  $T_{CDW}\simeq$135~K and then,
at zero field, a decrease  below $T_c$. The behaviour for $T < T_c$ suggested competition between the two types of order
with the CDW signal being restored when superconductivity was suppressed by a large magnetic field.
  However, the origin of the CDW ordering  vector, the driving force for  CDW order and the size of any CDW gap in the electronic density of states (DOS)  are still unknown.

One  fundamental thermodynamic property that could contribute towards understanding these questions  is the $T$-dependent static magnetic susceptibility $\chi(T)$
above $T_c$. Here we report measurements of $\chi_c(T)$ and
$\chi_{ab}(T)$ with magnetic field $H$ parallel and perpendicular to the crystalline
$c$-axis for relatively large twinned single crystals of YBa$_{2}$Cu$_{3}$O$_{6+x}$ at 7
values of the oxygen content $x$. In other CDW materials such as K$_{0.3}$MoO$_3$~\cite{bluebronze} the opening of the CDW gap causes a clear reduction in $\chi(T)$, and an increase in $d\chi(T)/dT$ for $T \leq T_{CDW}$. Using a special method of analysis, that will also be useful in the future for studies of detwinned YBa$_{2}$Cu$_{3}$O$_{6+x}$  crystals and other underdoped cuprates, we show that for most, but not all,  values of $x$,  the onset of the CDW has  a small but detectable effect on $d\chi_{ab}/dT$. From our viewpoint the physical reason for its small size is that as mentioned before~\cite{Cooper2014} the electronic entropy at $T_{CDW}$ has already been strongly suppressed because of the pseudogap. Or equivalently there are only short
Fermi arcs rather than a large Fermi surface so the electronic DOS has already been heavily reduced by the pseudogap. In particular for our UD68 crystal with $T_c$ = 67.9~K   we compare our results with expectations from  the $T$-dependent CDW intensity  measured for UD67~\cite{Chang}, bearing in mind that the size of the CDW gap will be proportional to the square root of the intensity~\cite{Grunerbook}.

While performing this analysis we again  found the surprising result,  briefly mentioned  previously~\cite{Kokanovic1}, that the small isotropic Curie ($C/T$) terms in $\chi(T)$ disappear quite abruptly above 200~K. Since then,  Biscaras et al.~\cite{Biscaras} have shown that fast cooling of polycrystalline YBa$_{2}$Cu$_{3}$O$_{6+x}$ from 400~K to below 300~K, increases $C$ and have  suggested that Cu$^{++}$ spins at the chain ends give  a significant contribution to $C$, so that the longer chains in slowly cooled samples give a smaller Curie term.  We discuss this picture and an alternative one in which the spins arise from localized states in the pseudogap~\cite{IslamZn}. Finally using model parameters obtained by fitting our  $d\chi_{ab}/dT$ data for UD68, we calculate the changes in electronic entropy, heat capacity and the free energy density caused by the CDW. We discuss the   importance of  making high resolution heat capacity measurements of a detwinned UD67 crystal in the future, preferably combined with structural and magnetic susceptibility studies on exactly the same sample.  From the present work we estimate the product of the free energy density and the CDW coherence volume in UD68  showing it to be $0.3 k_BT$ at 100~K and $0.7k_BT$ at 80~K. These numbers are lower limits and could be 2 or 3 times larger if the effect of the pseudogap contribution is handled differently. Nevertheless it underlines the probable importance of thermal fluctuations that were previously mentioned  in Ref.~\onlinecite{Wu2015}.
\section{II. EXPERIMENTAL DETAILS}
  A  YBa$_{2}$Cu$_{3}$O$_{6+x}$ crystal was grown by Dr.~K.~Iida~\cite{Kokanovic1} several years ago at the International Superconductivity Technology Center, Superconductivity Research Laboratory, Morioka, Iwate, Japan, using a modified crystal pulling method, often called the solute-rich, crystal-pulling  method~\cite{Shiohara1997}. This method is unusual in that there is a layer of  green phase Y$_2$BaCuO$_5$ at the bottom of the Y$_2$O$_3$ crucible below Ba-Cu-O liquid. In this situation, the liquid is always saturated with yttrium. Y$_2$BaCuO$_5$  is dense and always stays at the bottom of the crucible. In this work, as in Ref.~\onlinecite{Kokanovic1}, we measured pieces cut from a 1.5 gm piece of this crystal. We have no independent estimates of possible inclusions of other phases in these pieces apart from our susceptibility data. Above about 150~K, possible inclusions such as  Y$_2$BaCuO$_5$~\cite{211chi}  and BaCuO$_{2+x}$~\cite{112chi} will give  a Curie-Wiess  term $\chi(T)\sim 0.375/(T+\theta)$ emu/mole with $\theta \sim 50$ and -70 respectively. For several values of $x$ the Curie terms of our crystals were barely detectable. Taking the lowest value $C$=16 $\pm10$~10$^{-4}$emu-K/mole, Table~1, shows that there was at most 0.4$\pm0.25$ mole$\%$ of these two compounds. The level of possible inclusions of CuO can be estimated by comparing the peak in $d\chi/dT$  observed for CuO crystals~\cite{CuOchi} between 200 and 230~K with the noise level in our data, for example in Fig.~\ref{XRDcomp}. This gives a maximum of 2~mole$\%$~CuO.  The value of $T_c$ for an optimally doped piece of this crystal was 92.6~K~\cite{Kokanovic1}, compared with 94.3~K in Ref.~\onlinecite{Liang}. Strong in-plane scattering, for example from Zn/Cu substitution, suppresses $T_c$ by $\sim$13~K/at.$\%$ Zn~\cite{Zn}, so in our crystals the number of strong in-plane scattering centers is at most 0.13~at.$\%$.

 Field-warming magnetisation curves $M_c(T)$ at 10 Oe after zero-field cooling, shown later in the inset to Fig.~\ref{anisotropy} were extrapolated linearly to zero  in order to define the critical temperature $T_c$.  These  values of $T_c$  are listed in Table 1. Samples are labelled by these $T_c$ values rounded  to the nearest
integer, with prefixes UD denoting underdoped.   The $M_c(T)$ data show sharp  superconducting
transitions with widths  between the 50 and 90 $\%$ points ranging from 0.6 to 1~K for all $T_c$ values. $M_{ab}(T)$ curves measured in the same way gave the same values of $T_c$ but their widths were a factor of 2 larger, probably because of the smaller values of the lower critical field, $H_{c1}$, parallel to the CuO${_2}$ planes.
\begin{figure}
\includegraphics[width=7.0cm,keepaspectratio=true]{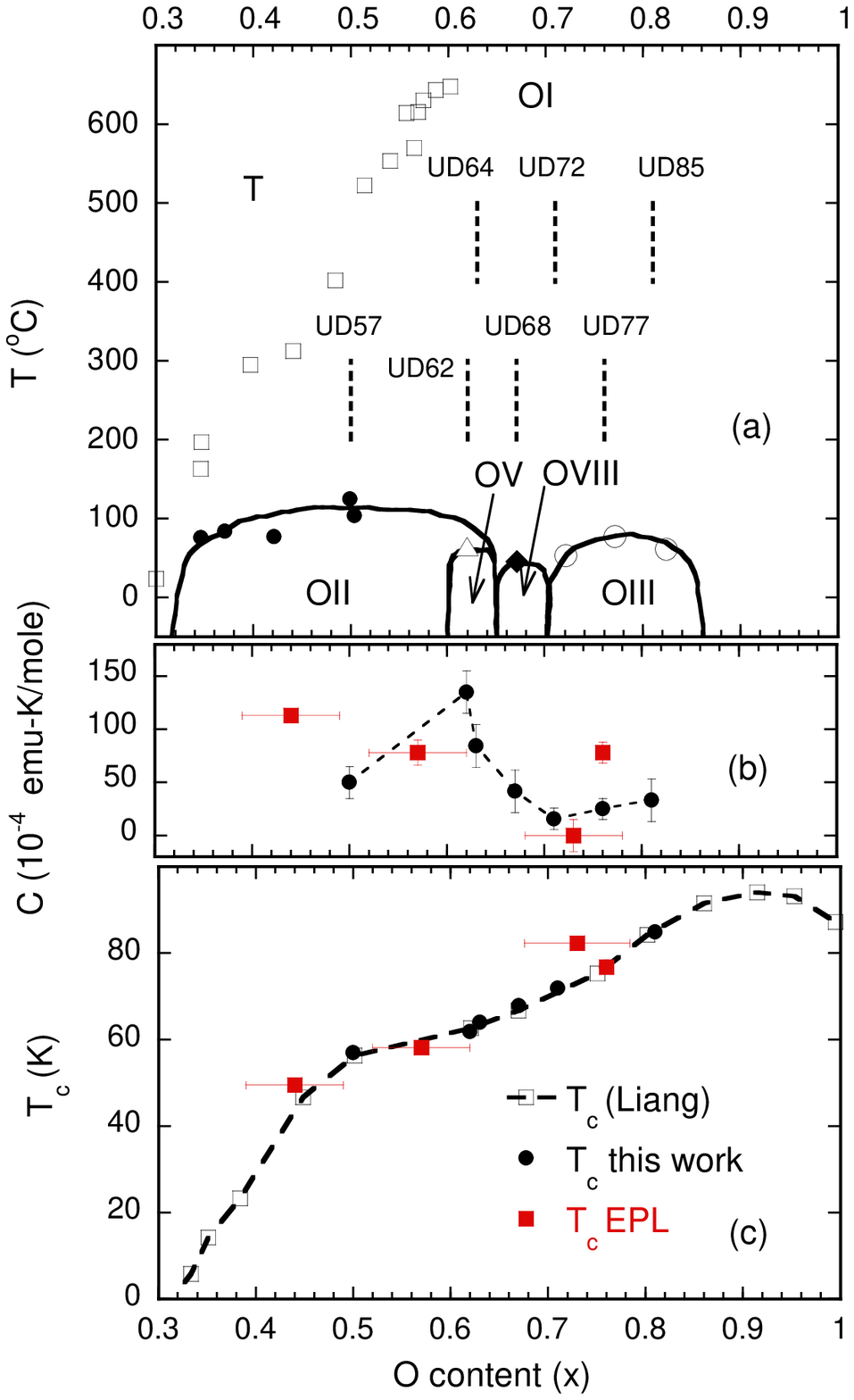}
\caption{ Color online. (a) Oxygen ordering diagram adapted from Ref.~\onlinecite{Zimmermann} showing the various ordered states of the Cu-O chains and the crystals studied here. (b) Curie terms for the samples reported here (black circles). (c) comparison of $T_c$  versus oxygen content $x$ in the present YBa$_{2}$Cu$_{3}$O$_{6+x}$ crystals after long annealing times between 20 and 25~C (black circles) with data for detwinned and annealed crystals from Ref.~\onlinecite{Liang}, open squares. In (b) and (c) data for annealing times of 1 to 2 hours from Ref.~\onlinecite{Kokanovic1} (red squares larger error bars) and for UD77 after 24 hours are also shown.}
 \label{Y123intro}
\end{figure}

The values of $x$ were determined more precisely than in our previous paper~\cite{Kokanovic1}.  Two separate pieces of the as-grown crystals, weighing 55 and 111 mg were annealed in flowing argon for  12 hours at 750~C, and their weight loss measured to 0.01 mg. A second identical anneal gave no further weight loss.   This procedure gave $x$ =  0.300$\pm0.004$, which is consistent with the observation that the samples were barely superconducting together with the $T_c(x)$ curve shown in Fig.~\ref{Y123intro}(c)~\cite{Liang}. Pieces of the as-grown crystal were then annealed under different conditions~\cite{Crystaldetails}, to give the seven samples UD85 through to UD57 with the values of  $x$ being obtained from the weight changes summarized in Table~1.
 In the present work  crystals  were mounted and
cooled in the SQUID magnetometer after being stored for 1 to 30 months at room temperature~\cite{Crystaldetails}. This is in contrast to our earlier study~\cite{Kokanovic1}, where SQUID measurements for the samples in Fig.~1 of Ref.~\onlinecite{Kokanovic1}, but not the other crystals in that paper, were made one to two hours after quenching the crystal on to a copper block. The oxygen contents of the crystals reported here are represented  by dashed lines on the oxygen ordering diagram in Fig.~\ref{Y123intro}(a) which is adapted from Ref.~\onlinecite{Zimmermann}. However we have no direct proof that this phase diagram applies to our crystals. They have  been annealed at room temperature for extremely long periods, during which weight changes were less than 0.01 mg, but they were not detwinned.     This weight change limit corresponds to an  uncertainty in $x$ ranging from 0.002 for a 234 mg crystal to 0.005 for an 83 mg one.  Because of this twinning it is important to note that as shown in Fig.~\ref{Y123intro}(c), $T_c(x)$  agrees with that of the
 crystals used for the CDW studies mentioned in the Introduction. For the CDW studies  the various ordered phases OV, OVIII etc. shown in Fig.~\ref{Y123intro}(a) were achieved using special annealing procedures applied to de-twinned crystals~\cite{LiangPhilMag}. However for some values of $x$ quantum oscillations were only  observed  for YBCO crystals where the OV and OVIII order had been suppressed by quenching from 100~C~\cite{Ramshaw} because surprisingly, at least for OVIII, this increases the carrier mean free path~\cite{Baglo}. Furthermore, structural studies on rapidly quenched crystals still show CDW effects although their amplitude is reduced~\cite{Achkarquench}. The annealing procedures should lead to longer Cu-O chains, an increase in $p$~\cite{Liang} and smaller Curie terms~\cite{Biscaras}.

  Fig.~\ref{Y123intro}(b) shows  the Curie constants $C$  for the present crystals together with three quenched crystals from  Ref.~\onlinecite{Kokanovic1}. The latter have larger uncertainties in $x$ which hampers detailed comparison. We can only say that the values of $C$ are similar. One further sample, UD77, for which OIII order could be expected, was measured one day after quenching and again 30 months later. There was no detectable change in $T_c$ but a substantial fall in $C$, as also shown in   Fig.~\ref{Y123intro}(b).  An
important caveat is raised later, but if we do  make the simple assumption~\cite{Biscaras} that each chain  end gives rise to one spin $s$ = 1/2, and assume that  a localized spin always separates two chain segments, then the number of spins is equal to the number of  chain segments. Hence the  values of $C$  in Fig.~\ref{Y123intro}(b) and Table~1 ranging from 135 to 16 10$^{-4}$emu-K/mole and corresponding 0.036 to 0.0043 $s$ = 1/2 spins per unit cell, give average chain segments ranging from 11 to 88 nm.  These are 2 to 15 times larger than CDW coherence lengths~\cite{Blanco}.

Optical microscope images of the twin patterns in our UD57, UD62, UD64 and UD68 crystals showed typical spacings of 1000~nm  corresponding to a limit of 1400~nm on the chain lengths from this source. The UD64 and UD68 crystals appeared to be completely twinned but for UD57 and UD62, twins could only be seen in about 50$\%$ of the area studied. For UD68 there were clear microcracks separated by $\simeq50\mu$m which aided oxygenation.   From the above arguments it probable that the twinning described here does not have significant effects on the CDW. However against  this, it is considered that twinning causes stress and macroscopic oxygen segregation in YBCO crystals~\cite{LiangPhilMag}. So similar measurements and analysis of detwinned YBa$_{2}$Cu$_{3}$O$_{6+x}$ crystals would be worthwhile if sufficiently large (50 mg or more) and uniformly oxygenated crystals can be obtained.
  \begin{figure}
\includegraphics[width=7.0cm,keepaspectratio=true]{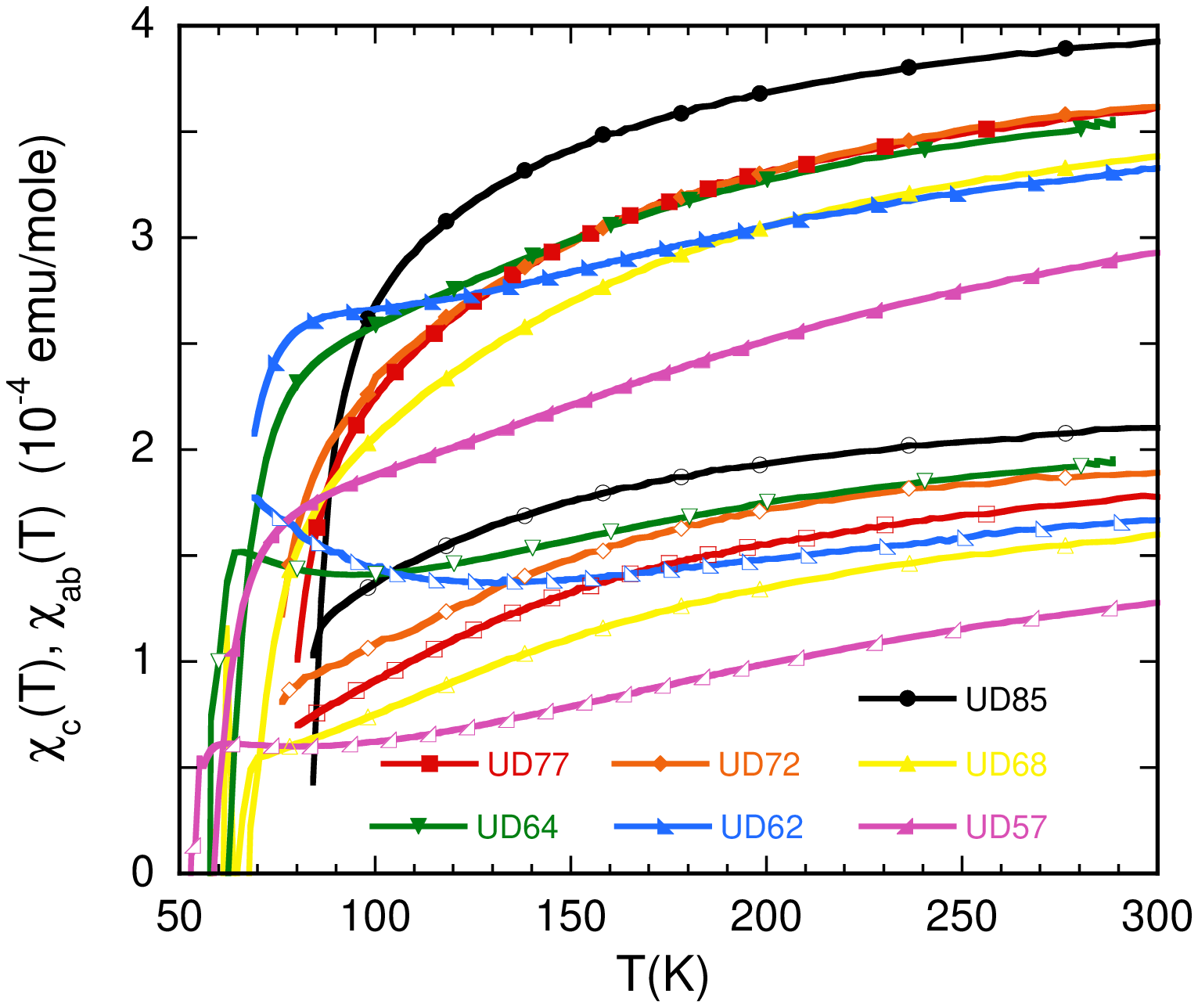}
\caption{ Color online.  Temperature dependence  of the static magnetic susceptibility
$\chi_{c}(T)$ (full symbols) and $\chi_{ab}(T)$ (empty symbols) for $H$ = 5 T $\parallel$ the $c$ axis and the $ab$
plane, respectively.}
 \label{Y123summary}
\end{figure}
\section{III. RESULTS AND ANALYSIS}
\begin{figure}
\includegraphics[width=7.0cm,keepaspectratio=true]{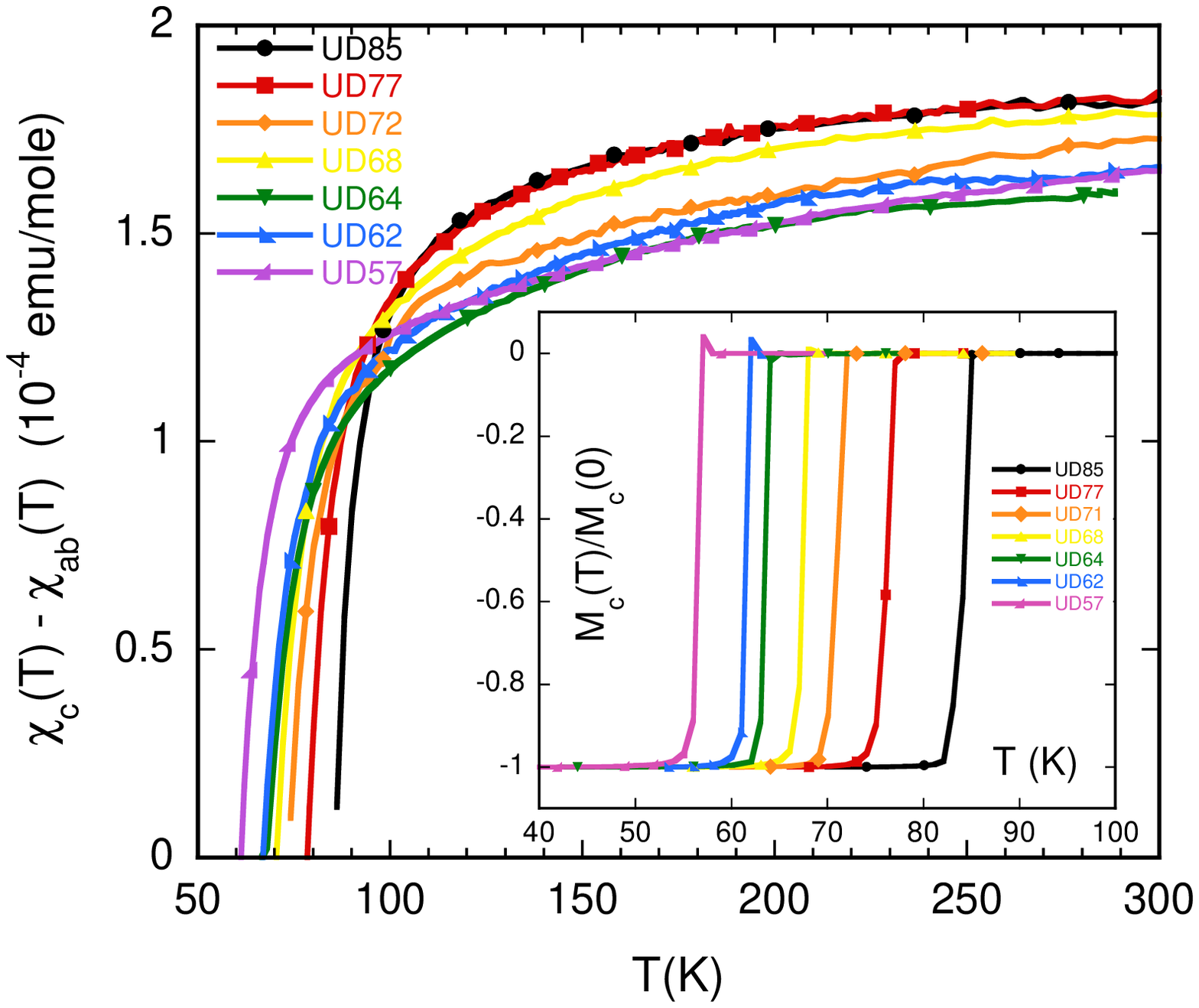}
\caption{Color online. Susceptibility anisotropy, $\chi_D(T)$=$\chi_c(T)-\chi_{ab}(T)$ for
YBa$_{2}$Cu$_{3}$O$_{6+x}$ single crystals. Inset: Superconducting transitions of the YBa$_{2}$Cu$_{3}$O$_{6+x}$ crystals for $H$ $\parallel c$ measured on warming in 10 Gauss after zero field cooling.}
 \label{anisotropy}
\end{figure}

The temperature dependences  of the static magnetic susceptibility
$\chi_{c}(T)$  $\parallel$ to the $c$ axis and $\chi_{ab}(T)$  $\parallel$ to the $ab$
plane for $H$ = 5 T  are shown in Fig.~\ref{Y123summary}, while Fig.~\ref{anisotropy} shows the susceptibility anisotropy, defined here as $\chi_{D}(T)$ = $\chi_{c}(T)$ - $\chi_{ab}(T)$, for the  seven YBa$_{2}$Cu$_{3}$O$_{6+x}$ crystals studied. It can be seen in Figs.~\ref{Y123summary} and \ref{anisotropy} that  above 200~K $\chi(T)$ increases with $T$ for all samples, which we 
ascribe to the pseudogap having a substantially larger value than any possible CDW gap, in line with arguments put forward in Refs.~\onlinecite{Cooper2014} and~\onlinecite{Briffa}. For UD64 and UD62 there are upturns below 150~K that are clearly visible in $\chi_{ab}(T)$  in Fig.~\ref{Y123summary} but are absent in the corresponding $\chi_{c}(T)$ - $\chi_{ab}(T)$ data in Fig.~\ref{anisotropy}.  This strongly suggests that there is an isotropic Curie term as found previously~\cite{Kokanovic1}.
\begin{figure}
\includegraphics[width=7.5cm,keepaspectratio=true]{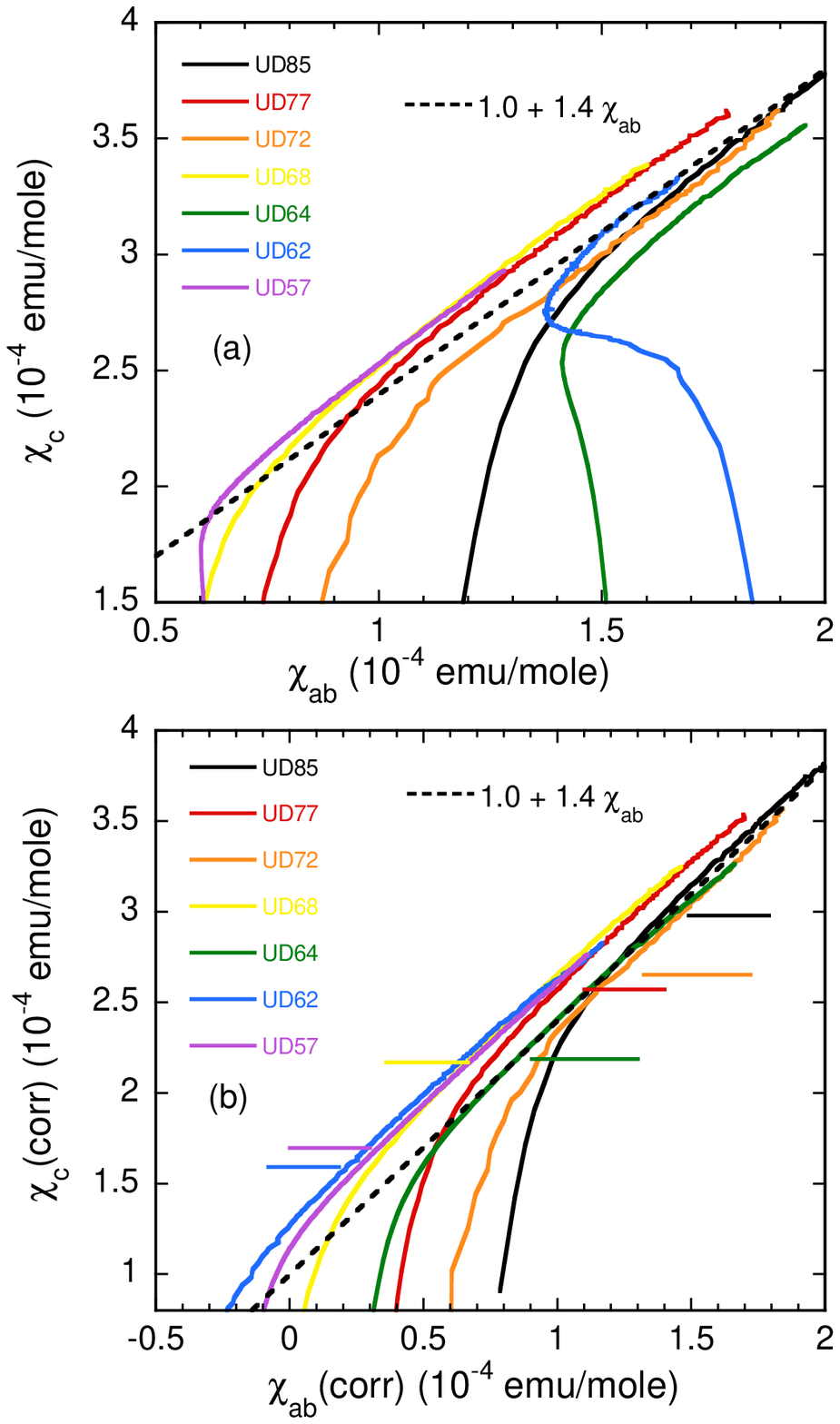}
\caption{Color online. Plots of static magnetic susceptibility
$\chi_{c}(T)$ with $H$ $\parallel$ the $c$ axis versus $\chi_{ab}(T)$ with $H \parallel$ to the $ab$ plane (a) for the data shown in Fig.~\ref{Y123summary} and (b) after subtracting the Curie terms shown in Fig.~\ref{Y123intro}(b) and Table~1. The dashed line shows the behavior expected for a constant anisotropic $g$ factor with  $(g_c/g_{ab})^2$ = 1.4~\cite{Kokanovic1}, the color-coded horizontal lines show $\chi_{c}(130)$.}
\label{chicvschiab}
\end{figure}
\subsection{A. Analysis of Curie terms}
\begin{figure}
\includegraphics[width=7.5cm,keepaspectratio=true]{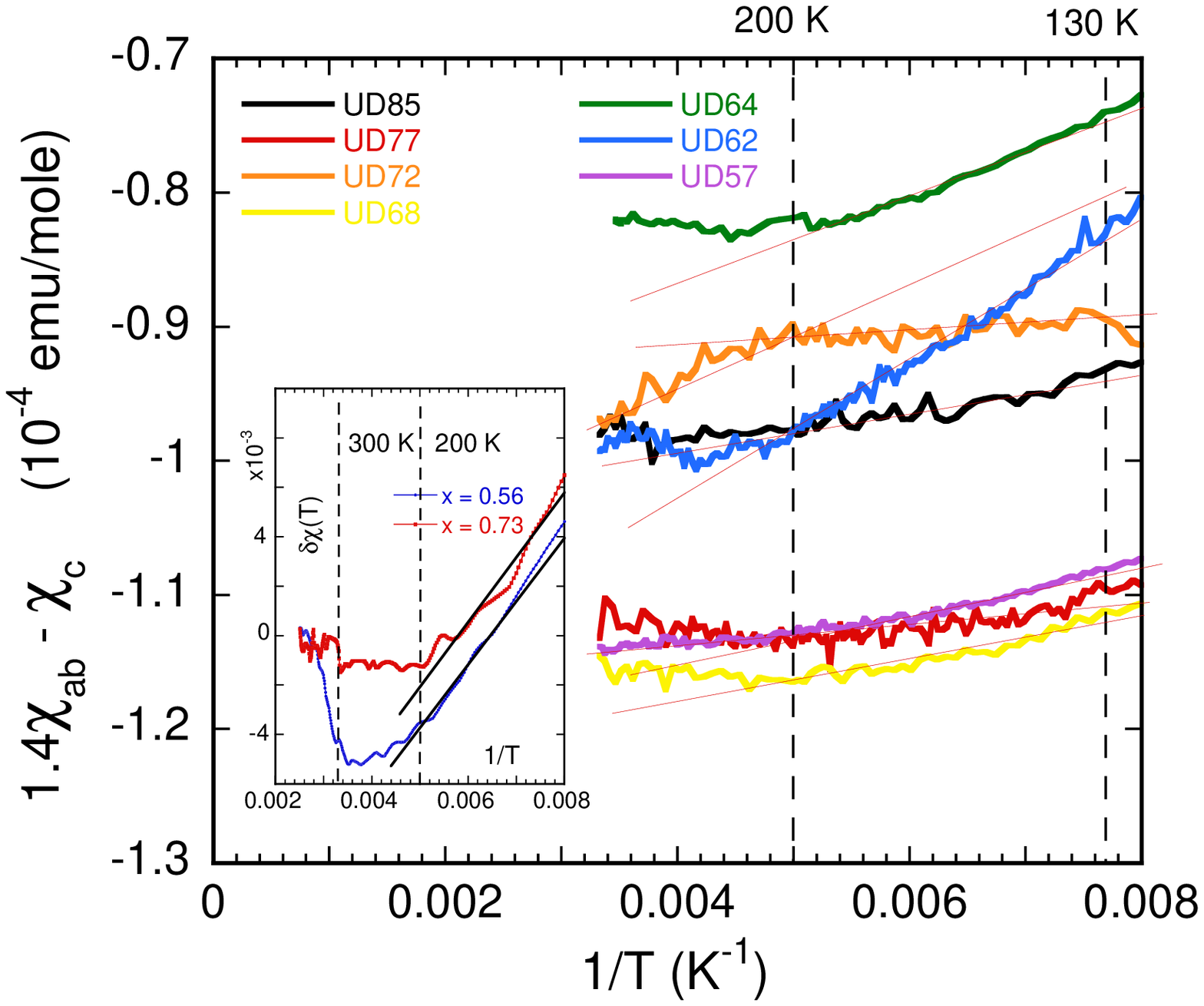}
\caption{Color online. Plots of 1.4$\chi_{ab}(T)$-$\chi_{c}(T)$ vs. $1/T$ for the long-time annealed crystals studied here. The linear regions between 200 and 130~K are used to determine the Curie terms, $C$ for each sample, shown in Fig.~\ref{Y123intro}(b) and in the inset to Fig.~\ref{chiabcorr}. Errors in $C$ were obtained by making similar plots (not shown) for $(g_c/g_{ab})^2$ = 1.35 and 1.45. The inset shows data from Ref.~\onlinecite{Biscaras} where $\delta\chi(T)$ (converted to the units used here) is the difference between fast and slowly cooled polycrystalline samples of YBCO with the $x$ values shown. In all cases, except UD72(see text), the Curie term tends to go away above 200~K.}
\label{Curie}
\end{figure}

Fig.~\ref{chicvschiab}(a) shows plots of $\chi_{c}(T)$ vs.~$\chi_{ab}(T)$ for all samples studied with $T$ as an implicit parameter. This method of analysis was originally applied to Bi$_2$Sr$_2$CaCu$_2$O$_{8+\delta}$ crystals~\cite{Watanabe}. With the notable exceptions of  UD64 and UD62 they are linear with a slope of 1.4 over an extended range before turning down because of the diamagnetic superconducting fluctuation term which is predominantly in $\chi_{c}(T)$ because of the large anisotropy~\cite{KokanovicR}. This downturn sets in quite abruptly somewhat below $\simeq$130~K. Therefore in Fig.~\ref{Curie} we plot 1.4$\chi_{ab}(T)$-$\chi_{c}(T)$ vs. $1/T$ down to 125~K. Such a plot eliminates any susceptibility contributions arising from the pseudogap as discussed previously~\cite{Kokanovic1} and from any CDW under the reasonable assumption that it also has a $g^2$ anisotropy of 1.4  as suggested by the dashed line in Fig.~\ref{chicvschiab}(a). It therefore allows the isotropic Curie term to be identified.
The linear regions in Fig.~\ref{Curie}  between 200 and 130~K give the Curie constants for each crystal. When they are subtracted from both $\chi_{c}(T)$ and $\chi_{ab}(T)$ the corrected plots are then linear down to at least 130~K as shown in Fig.~\ref{chicvschiab}(b) where the values of $\chi_{c}(130)$ are marked by horizontal lines.
 A surprising result in Fig.~\ref{Curie} is that the Curie plots for all crystals flatten off and deviate from $1/T$ behavior above 200~K.
The inset to Fig.~\ref{Curie} shows Curie plots obtained by scanning in data from Figs.~1 and 3 of Biscaras et al~\cite{Biscaras} and converting them to the units used here. These refer to experiments in which the same polycrystalline samples of YBa$_{2}$Cu$_{3}$O$_{6+x}$ were measured using a SQUID magnetometer in both the rapidly cooled (10~K/min) and slowly cooled (1~K/min) states. Fast cooling gives less order in the  Cu-O chains below 300~K and hence slightly larger Curie terms that are visible in the difference plots shown.   The data for $x$ = 0.73   shown in the  inset to Fig.~\ref{Curie} are particularly clear because there are large deviations from a Curie law above 200~K but only minor changes at higher $T$. This strongly supports our finding that the Curie plots  flatten off above 200~K. However the magnitudes of the  Curie terms obtained from the main part of Fig.~\ref{Curie}, that are tabulated in Table~1, are typically a factor of 10 larger than the value for difference plots, $C$= 2.7 10$^{-4}$emu-K/mole shown by the straight lines in the insert to Fig.~\ref{Curie}.

Because all our samples are derived from the same large crystal we can confirm the result of Ref.~\onlinecite{Biscaras} that the Curie terms are caused by oxygen disorder in the chains and therefore faster cooling increases $C$ by $\sim10\%$. The results in Fig.~\ref{Y123intro} (b) suggest that this disorder could be particularly large for OV crystals and that for OIII ones  long annealing times at room temperature increases the chain order.  However we feel that further evidence  such as ESR or NMR data is  needed in order to be
sure that the $C/T$ terms are indeed caused by Cu$^{++}$ ions in the chains.  For Zn doped La$_{2-x}$Sr$_x$Cu$_{1-y}$Zn$_y$O$_4$ samples it  was found that Zn substitution introduced Curie-like behavior~\cite{IslamZn}. In this case it  was argued~\cite{IslamZn} that  disorder induces  a transfer of spectral weight from the wings of the pseudogap to low energies. It is possible that oxygen disorder in the Cu-O chains of YBCO is having the same effect. Some support for this picture is given by  Curie plots (not shown here) of the difference data in Ref.~\onlinecite{Biscaras}. For $x$  values ranging from 0.79 to 0.43, but not for lower values,  departures from $C/T$ behavior occur at gradually increasing temperatures as $x$ decreases, i.e. as the pseudogap gets larger.

  In either of the scenarios involving oxygen disorder in  the chains, the flattening out above 200~K  could be connected with evidence from thermal expansion studies~\cite{Meingast}  for time-dependent ``glassy'' behavior down to 230~K for YBa$_{2}$Cu$_{3}$O$_{6.95}$. This is absent for YBa$_{2}$Cu$_{3}$O$_{7.0}$ where the Cu-O chains are all full. The abrupt change in $\delta \chi$ above 300~K  for $x=0.56$ shown in the inset to  Fig.~\ref{Curie} is caused by a decrease in $p$ when the slowly cooled sample is heated above 300~K.  At this point the Cu-O chains in slowly cooled samples become less ordered and $p$ falls slightly, as noted previously~\cite{Walker} in work on the electrical resistivity and the Hall effect of epitaxial YBCO films.
\begin{figure}
\includegraphics[width=7.0cm,keepaspectratio=true]{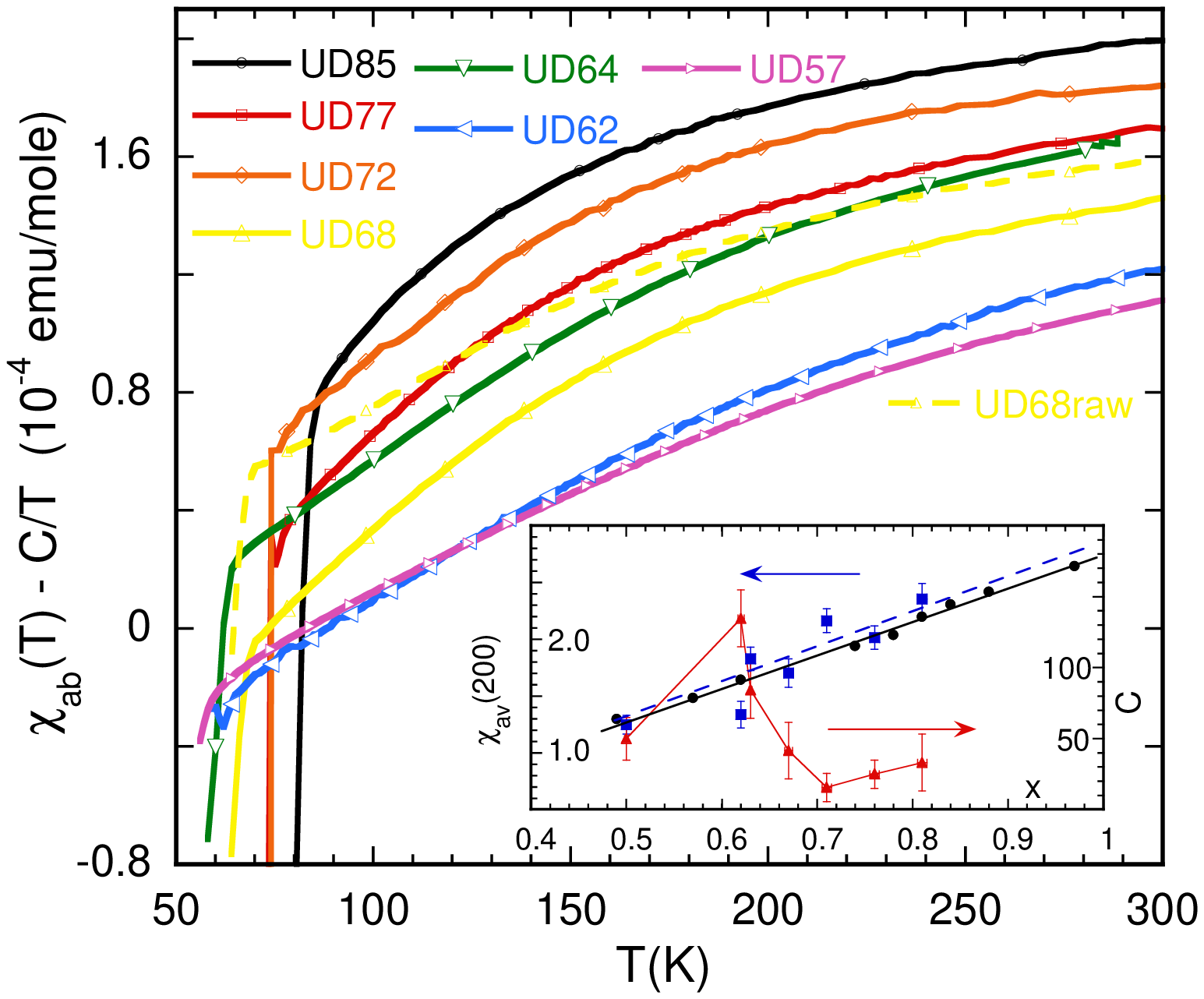}
\caption{Color online. In-plane magnetic susceptibility $\chi_{ab}(T)-C/T$ of the 7 YBa$_{2}$Cu$_{3}$O$_{6+x}$ samples studied after subtraction  of the Curie terms $C/T$ obtained using the procedure given in the text. For UD68 the measured values of $\chi_{ab}(T)$ are also shown.  The inset shows the dependence of three quantities on oxygen content $x$:  the average susceptibility $\chi_{av}\equiv \frac{2}{3}\chi_{ab}+\frac{1}{3}\chi_{c}$ at 200~K (blue squares) after subtracting the Curie terms, published  data for polycrystalline YBa$_{2}$Cu$_{3}$O$_{6+x}$~\cite{Cooper1996}, black circles, where no Curie term has been subtracted and the Curie term $C$ itself in units of $10^{-4}$emu-K/mole. $C$= 37.5$\times 10^{-4}$emu-K/mole  corresponds to  0.01 spin $\frac{1}{2}$ centres per formula unit.}
\label{chiabcorr}
\end{figure}
Fig.~\ref{chiabcorr} shows the  values of $\chi_{ab}(T)$ for the 7 YBa$_{2}$Cu$_{3}$O$_{6+x}$ samples studied, after correction for the Curie terms. These do not have any contributions from localized spins or from superconducting fluctuations~\cite{KokanovicR} (except possibly a few K above $T_c$) and  can therefore be examined for the possible influence of CDWs. It is clear that $\chi_{ab}(T)$ is continuing to rise at 300~K, which we ascribe to a separate effect from the pseudogap in agreement with several other research groups~\cite{Wu2015,Cooper2014,Briffa}. Apart from  the $T$ dependence caused by the pseudogap, there is no  obvious sign of CDW effects with an onset temperature $T_{CDW} \simeq$150~K found in the NMR~\cite{Wu2015} and structural~\cite{Achkar,Blackburn,Chang,Blanco} studies cited in the Introduction. We therefore take a closer look at $d\chi_{ab}(T)/dT$ in the following Section.
\begin{figure}
\includegraphics[width=6.5cm,keepaspectratio=true]{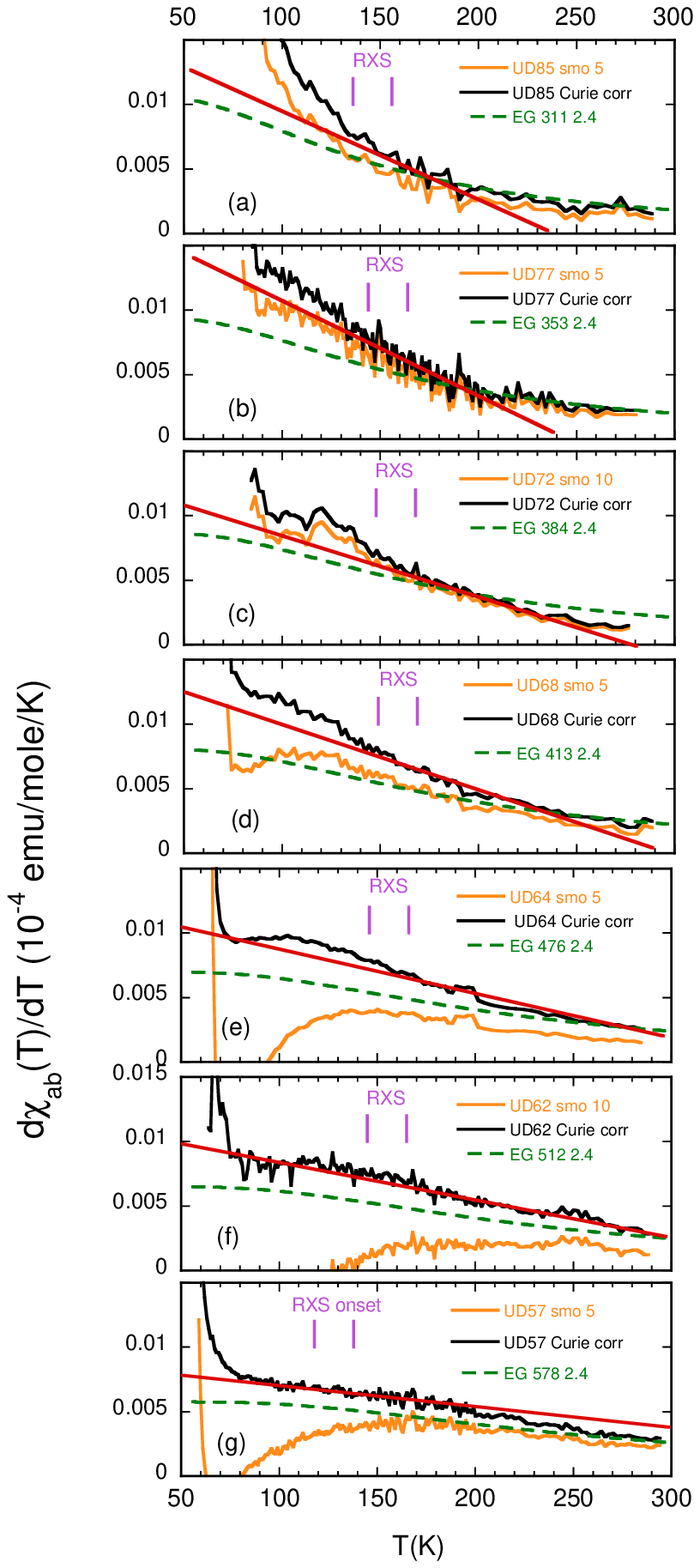}
\caption{Color online.  Smoothed derivative plots $d\chi_{ab}(T)/dT$ vs. $T$ for all crystals studied here, orange lines.   Solid black lines show the effect of adding $C/T^2$ which removes the Curie term arising from localized spins.   The  red lines and the green dashed lines show different ways of defining the background signal caused by the pseudogap (see text). The  $T_{CDW}$ onset temperatures, including their errors bars, obtained from resonant X-ray scattering for each value of $x$~\cite{Blanco} are shown by the vertical purple lines.}
\label{chiabderiv}
\end{figure}
The $T$-dependent curves in Fig.~\ref{chiabcorr} would be appropriate for comparison with local probes of the spin susceptibility such as Y$^{89}$ Knight shifts~\cite{AlloulRev}  or Gd$^{3+}$ electron spin resonance shifts~\cite{JanossyESR}, although the region above 200~K should be handled with care because in Fig.~\ref{chiabcorr} we have subtracted the $C/T$ Curie correction term over the full range of $T$ even though we think that the localized spins become delocalized above 200~K. The variation of these curves with oxygen content $x$ is reasonably smooth and monotonic after including possible measurement errors and the errors in $C$. This is illustrated  in the inset to Fig.~\ref{chiabcorr} where the average susceptibility $\chi_{av}\equiv\frac{2}{3}\chi_{ab}+\frac{1}{3}\chi_{c}$ at 200~K (where we can be more confident of the Curie term $C$ and are still well above the  superconducting fluctuation region) is compared with the smoother variation obtained for polycrystalline samples~\cite{Cooper1996}. After including conservative measurement errors (corresponding to the absence or presence of 2 mm of the plastic tube holding the sample) and the errors in $C$, the data for all samples except UD62 are reasonably consistent with the blue trend line  for  crystals and the black line for polycrystalline samples shown in the inset.  The blue line is drawn slightly higher than the black one  because  a small amount of  preferential alignment in our polycrystalline samples reduced $\chi_{av}$ slightly~\cite{Kokanovic1}.   We note that the extrapolated intercepts in Fig.~\ref{Curie} at $1/T$ = 0 have a spread of 0.3 10$^{-4}$ emu/mole. If this is converted into  a standard deviation ($\sigma$) in $\chi_c$ and $\chi_{ab}$ (taken to be the same) it gives $\sigma$ = 10$^{-5}$ emu/mole, consistent with the scatter and the error bars in the inset to Fig.~\ref{chiabcorr} and with the level of non-monotonic behavior with $x$ that can be seen in the main part of this figure.
The exception to the above  discussion is UD62 for which $C$ is 135$\pm20$ 10$^{-4}$ emu-K/mole or 85$\pm25$ 10$^{-4}$ emu-K/mole larger than that for UD58.  In  Ref.~\onlinecite{Biscaras} it was argued that for $x$ = 0.45 one spin 1/2 centre per unit cell decreases the planar hole concentration by 0.5 per unit cell. Applying the same model to UD62 reduces its  $p$ value to 0.103$\pm0.002$ (relative to UD58) which is very close to that for UD58, in line with their values of $\chi_{ab}(T)-C/T$ and $\chi_{av}(200)$ in the main part of Fig.~\ref{chiabcorr} and the inset respectively.
  In summary, from the discussion in this Section we conclude that the Curie term in our samples is  important for detailed understanding and that it is reasonably well understood at an empirical level.
\subsection{B. Possible effects of  CDWs on derivative plots}
Fig.~\ref{chiabderiv} shows derivative plots, $d\chi_{ab}(T)/dT$ vs. $T$ for the 7  long-time annealed crystals studied here  on the same scales,  with a view to detecting the onsets of CDW order in the temperature ranges given by RXS~\cite{Blanco}. Raw derivatives (not shown) were obtained  by subtracting neighbouring $\chi_{ab}(T)$ points and dividing by the 1 to 4~K difference in their $T$ values. They were then smoothed using a sliding average of 5 points. We have also calculated the derivatives from the raw data using sliding second order polynomial fits to groups of 3, 5 or 7 adjacent points. These give the same results with no significant differences for different smoothing  ranges.
The orange curves show the derivatives after smoothing while the black curves include the  $C/T^2$ correction for the Curie term. Note that this has been added for all $T$ even though Fig.~\ref{Curie}  shows clear deviations from a Curie law above 200~K. For several values of $x$, $C$ is small so there is little difference between the orange and black curves. For the samples where $C$ is larger, there is no obvious transition between the black and orange curves above 200~K. We believe the flattening of the Curie plots above 200~K is largely  compensated either by a small increase in $p$ above 200~K, in the ``Cu$^{++}$ chain scenario'', or by an effective reduction of the pseudogap energy in the `` low energy spectral weight'' picture.

In order to see the effect of the CDWs we have to define a $T$-dependent background caused by the pseudogap. In view of the flattening   of the Curie plots  above 200~K we have defined our ``best'' background as a straight line that fits the data from the onset T$_{CDW}\sim$~150~K given by RXS up to 200~K.  These straight lines are shown in red in Fig.~\ref{chiabderiv}.
An alternative way of determining the background  contribution to $d\chi_{ab}(T)/dT$ is to use Loram's triangular gap model for the DOS.
This is a  first-order phenomenological model, describing a nodal pseudogap, that accounts for a large body of heat capacity (entropy) and magnetic susceptibility data~\cite{Loram01} for  polycrystalline cuprates and some magnetic susceptibility data for single crystals~\cite{Kokanovic1}. It gives:
\begin{equation} \chi_{PG}\left(T\right) = A\left\{1-y^{-1}\mathrm{ln}\left[\mathrm{cosh}y\right]\right\}
\label{eqn:chipg}
\end{equation}
where  $A= N_{0}(g\mu_{B})^{2}$, $y={E_G}/{2k_{B}T}$, $N_{0}$ is the electronic DOS at high energies, assumed to be independent of energy, and $E_G$ is the pseudogap energy. Previously~\cite{Kokanovic1} we took $A$ = 3.0 and 4.2$\times10^{-4}$ emu/mole for $\chi_{ab}$ and $\chi_{c}$ corresponding to an average spin susceptibility at high $T$ of 3.4$\times 10^{-4}$ emu/mole.  This model does not include Fermi arcs which according to  Ref.~\onlinecite{Franz} are ``protected'' regions that are responsible for superconductivity in under-doped cuprates.  The length of these  arcs can  be estimated  from the ratio of the superfluid densities (squares of the London penetration depths) at low $T$ for YBa$_{2}$Cu$_{3}$O$_{7}$ and YBa$_{2}$Cu$_{3}$O$_{6.5}$~\cite{Pereg}. In a minimal model with an underlying  cylindrical Fermi surface the arcs are  $\simeq 20\%$ of the circumference.

Provided the pseudogap energy increases linearly from zero at the tips of the arcs towards the anti-nodes, the triangular gap model will still apply  but with  $A$ values reduced by a factor 0.8 and with an additional $T$-independent paramagnetic contribution of 0.2$A$, which will not show up in the derivative. The green dashed lines in Fig.~\ref{chiabderiv} show calculated $d\chi_{ab}(T)/dT$ vs. $T$  curves obtained from Eqn.~\ref{eqn:chipg} with these reduced values of $A$ and $E_G/k_B$  given by 1200[1-(p/0.19)]~\cite{Loram01}, with values of $p$ given in Table~1 that were determined from the $T_c(p)$ relation of  Liang et al.~\cite{Liang}.

With the exception of UD72 (which may be atypical in that there appears to be a Curie term above 200~K in Fig.~\ref{Curie} and none below) and UD62 which has a relatively large Curie term, the green curves give  reasonably good fits to the data in Fig.~\ref{chiabderiv}, but as shown later for UD68 on an expanded $y$-scale in Fig.~\ref{XRDcomp} they are not as good as a linear background. They also tend to give $T_{CDW}$ values that are somewhat higher than RXS, but this would be consistent with Raman studies where $T_{CDW}\simeq$ 175-200~K \cite{Raman}. For the green curves,  CDW effects in the derivative plots near 100~K  are up to a factor of 2 larger than when using the red lines as background. However we note that taking $A$ = 2.4$\times10^{-4}$ emu/mole  is based on the assumption of a cylindrical Fermi surface. Band-structure calculations usually give a larger DOS in the antinodal directions~\cite{Eschrig2006}. Including this effect would improve the fits, since it would require a larger value of $A$. We therefore focus  on the red lines and use the green curves to estimate an upper error bound. From the difference between the black curves and the red lines in Fig.~\ref{chiabderiv}  we see the effect of the CDW at 100~K  for UD85 through to UD68 is approximately the same, for UD64 it is a factor of 2 smaller while for UD62 and UD57 it is hardly detectable. We note that the  deviations from the $T_c(p)$ parabolic law are still substantial for UD85 samples~\cite{Liang}.  Therefore bearing in mind our argument that the $p$ values for UD57 and UD62 are essentially the same, we can conclude that all the data in Fig.~\ref{chiabderiv} is consistent with the suggestions, previously formulated in terms of electronic phase separation~\cite{Tallon1997} and stripe order~\cite{Liang,AndoHall,Taillefer}, that the CDW actually causes the plateau in $T_c(x)$ by suppressing $T_c$ below the parabolic law when $p$ is between 0.11 and 0.14.
\subsection{C. Numerical comparison with hard X-ray diffraction data for UD67.}
\begin{figure}
\includegraphics[width=7.5cm,keepaspectratio=true]{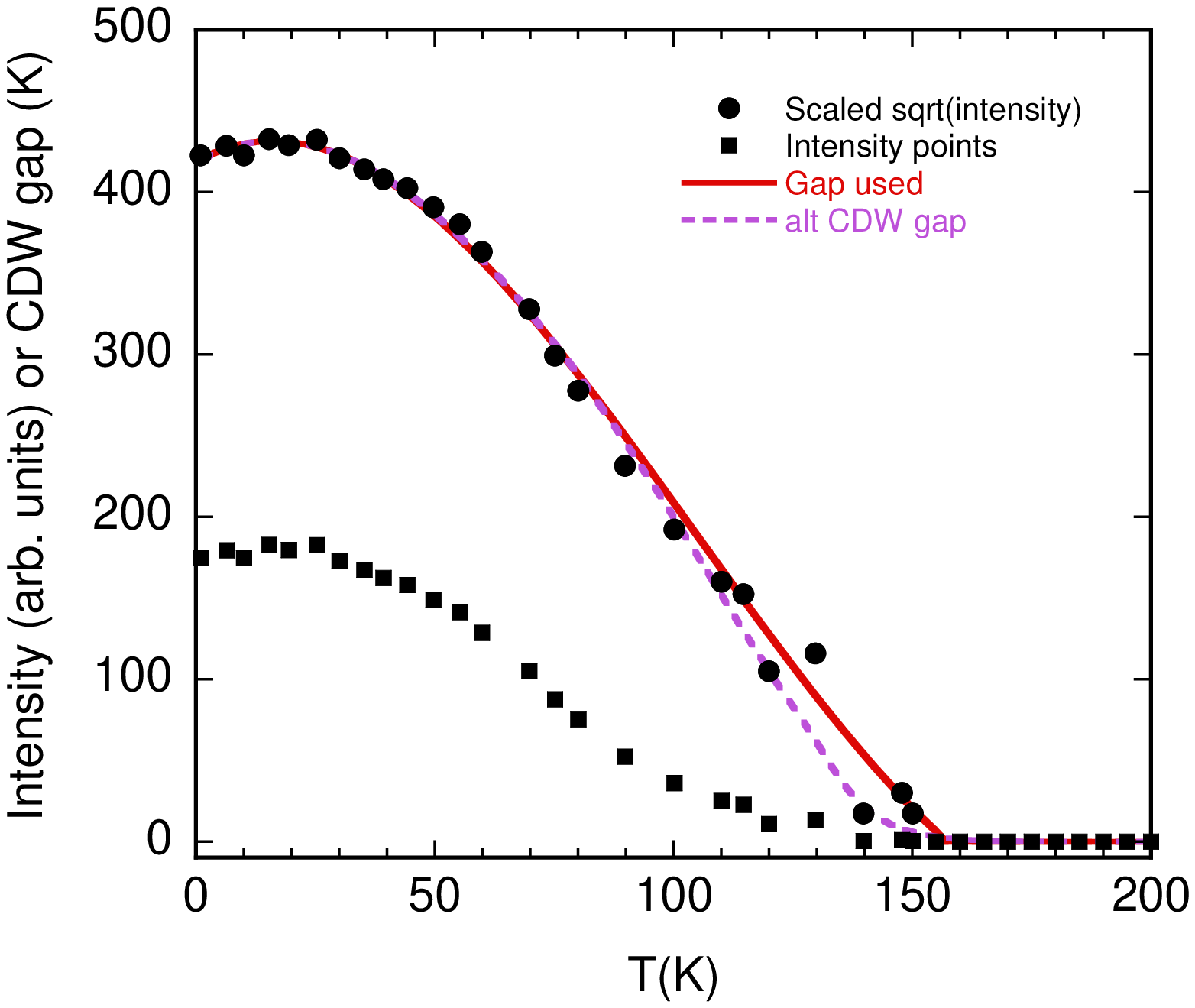}
\caption{Color online.  Intensity (black squares) of CDW diffraction peak vs.$T$ for UD67  in an applied field of 17~T, that severely weakens superconductivity,  taken from Ref.~\onlinecite{Chang} and the corresponding CDW energy gap (solid circles), assuming that it has a maximum value of 433~K or 37.3 meV. Note the relatively small values, $\Delta_{CDW}(T)/k_BT<2$ above 100~K. The solid red curve shows the $T$-dependent CDW gap used in model calculations. The alternative dashed curve gives slightly smaller effects in $d\chi_{ab}(T)/dT$  just below $T_{CDW}$.}
\label{XRDandgap}
\end{figure}

In order to estimate the magnitude of the changes expected in $d\chi_{ab}(T)/dT$ we use the published hard X-ray diffraction data of Chang et al.~\cite{Chang} for UD67 and make a comparison with our data for UD68 that has $T_c$ = 67.9~K.  Any CDW gap, $\Delta_{CDW}(T)$, in the electronic DOS will be proportional to the amplitude of ionic displacements~\cite{Grunerbook} that is to  the square root of the intensity of the CDW diffraction peaks. In view of the competition between CDWs and superconductivity~\cite{Chang} it is reasonable to assume that $\Delta_{CDW}(0)\simeq$ 36~meV, a typical value quoted for the maximum of the $d-$wave superconducting energy gap parameter in under-doped cuprates~\cite{Tunnellingreview}. This normalization gives $\Delta_{CDW}(T)$ shown in Fig.~\ref{XRDandgap}.

We use a  model in which there is a uniform CDW gap, of the same states-conserving form that also occurs in superconductivity, over a certain length of the Fermi arcs (specified by the parameter ``fract'' in Fig.~\ref{XRDcomp}) and then calculate $\chi_{ab}(T)$ and $d\chi_{ab}(T)/dT$ using standard expressions and Fermi statistics.
 It can be seen that the magnitude and $T$ dependence of the effect are well described with $\Delta_{CDW}(0)$ = 433~K and the CDW gap only extending over 25$\%$ of the length of the arcs, i.e. over only  5$\%$ of the circumference of the large quasi-two dimensional Fermi surface present in over-doped cuprates. Even for such a large value of $\Delta_{CDW}(0)$, $\Delta_{CDW}(T)$ above 100~K is comparable with, or less than 2$k_BT$ so one would not expect to see  effects associated with closed  electron  pockets on transport properties in low magnetic fields in this range of $T$. This provides support for the model involving electron-electron scattering between Fermi arcs  proposed by Gor'kov~\cite{Gorkov} in connection with the $T^2$-dependence of the in-plane resistivity of UD HgBa$_2$CuO$_{4+x}$~\cite{Barisic}.

 On the other hand there are indications of  possible inconsistencies with other work  that need further study. For a $d$-wave superconducting gap that rises to  433~K at the anti-nodes, then for arc lengths of 20$\%$ the gap at the ends of the arcs, will only be 433~$\sin(18)$ or 134~K which would not be large enough to suppress a CDW with a gap of 433~K. As shown in Fig.~\ref{XRDcomp} a smaller value of $\Delta_{CDW}(0)$ = 216~K does not give a good fit to the data. The parameter ``fract'' has to be  1.0 in order to account for the slope of the data between 120 and 150~K. and then this value of $\Delta_{CDW}(0)$ does not give such a good fit at lower $T$, as shown in Fig.~\ref{XRDcomp}.  ARPES data for several cuprates (including Bi$_2$Sr$_2$CuO$_{6+\delta}$) shown in Fig.~10(b) of Ref.~\onlinecite{Yoshida2012} give average arc lengths just above $T_c$ rising  from 23 to 27$\%$  between $p$ = 0.1 and 0.125 which is reasonably consistent with our minimal model. For some unknown reason the arc length for Bi$_2$Sr$_2$CuO$_{6+\delta}$ at $p$ = 0.12 shown in Ref.~\onlinecite{Comin} is much larger (50$\%$). We should also mention that in the Fermi arcs picture used here there is nothing particularly special about the doping level $p=1/8$ where the suppression of $T_c$ by the CDW reaches a maximum~\cite{Liang}.

 An alternative picture which would be in better agreement with scanning tunnelling spectroscopy studies~\cite{STSarcs} would be to have longer arcs. In this case the arcs would need to have lower spectral weight in order to maintain the same superfluid density and the scenario in which the CDW wave-vector connects the tips of the arcs would be less convincing. In this picture  the detailed model is somewhat different but any CDW-induced anomalies in $d\chi_{ab}(T)/dT$ would still be small, because of the reduction in electronic DOS and entropy caused by the reduced spectral weight.

 However we are reluctant to abandon the  concept of Fermi arcs because it could be very important for understanding the effect of a magnetic field on the CDW. The Lorentz force from a magnetic field perpendicular to the CuO$_2$ planes  drives electrons from one end of the Fermi arc to the other. In a very simple qualitative  picture this tends to pile up charge at one tip of an arc and to deplete it at the other, which would enhance a CDW in an applied magnetic field even when superconductivity has been heavily weakened or destroyed. There is experimental evidence in favor of this suggestion in that the CDW intensity in UD67 increases strongly between  25 and 28~T~\cite{Gerber}.
\subsection{D. Estimates of the changes in free energy associated with the CDW transitions.}
The small size of the CDW anomalies discussed in the previous Section raises  questions as to what is driving the CDW and how much free energy is gained by opening a CDW gap in the electronic DOS. We can answer the second question by continuing to model the data in terms of  fermionic quasi-particles.   Although this is the electronic contribution, in the harmonic approximation there is no change in the lattice entropy in the CDW state,~\cite{Tutisprivatecomm} so this  a reasonable initial estimate. Despite the undoubted importance of electron-electron correlations, a fermion picture seems be relevant for  hole-doped cuprates
 because their Wilson ratio is close to the value for weakly interacting fermions~\cite{Cooper1996}.

\begin{figure}
\includegraphics[width=7.0cm,keepaspectratio=true]{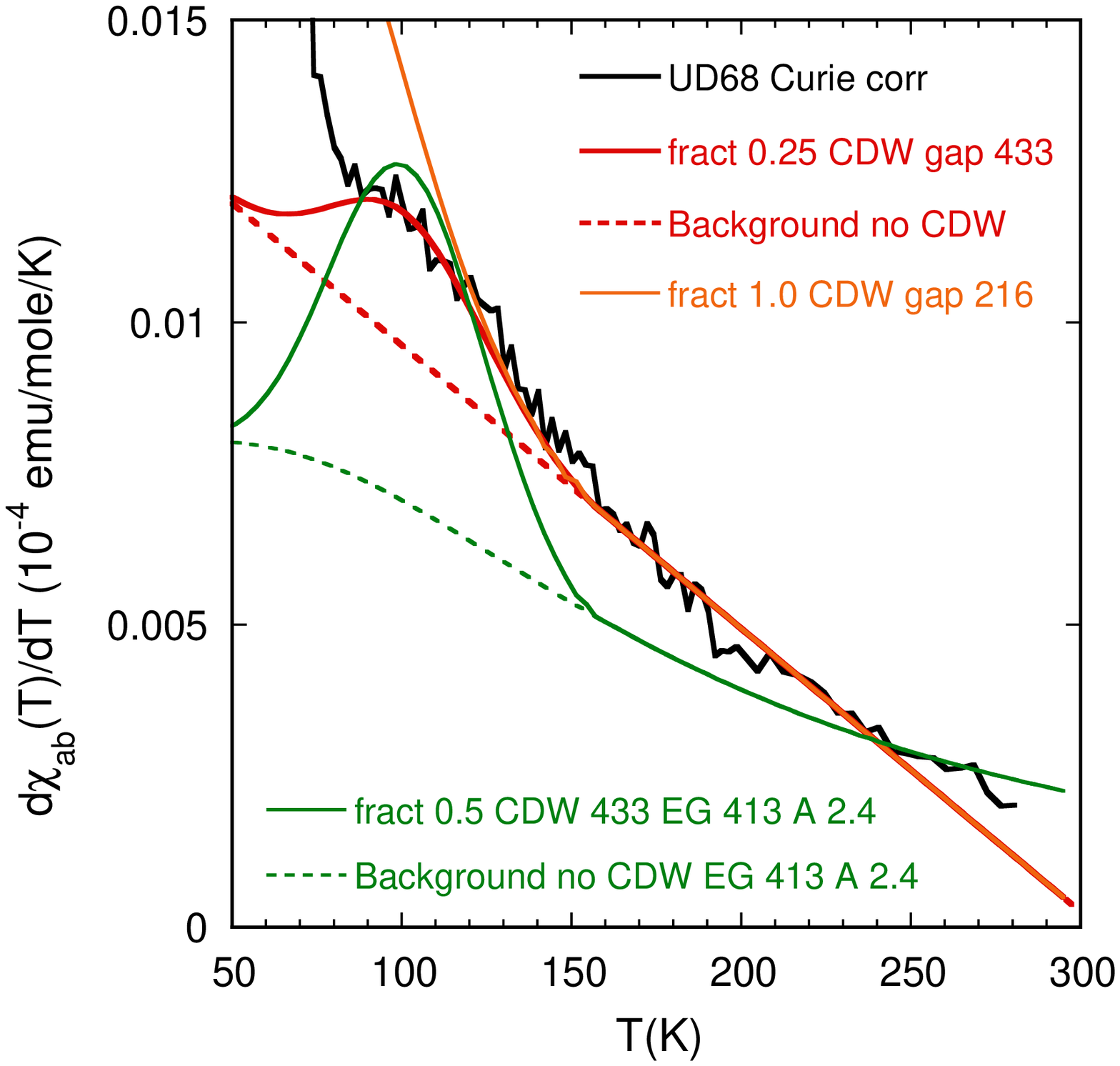}
\caption{Color online. Comparison of measured values of $d\chi_{ab}(T)/dT$ for UD68($T_c$ = 67.9~K), with calculations using CDW energy gaps derived from the X-ray intensity shown in Fig.~\ref{XRDandgap} and a simple Fermi arc, solid red curve, and pseudogap model, green curve (see text). The best agreement, shown by the solid red line, is given for a CDW gap  rising to $\sim433$~K at low $T$ that reduces the lengths of the arcs by a factor $\sim0.75$. Deviations from the solid red line below 80~K could arise from a reduction in electronic DOS caused by superconducting fluctuations. Green curves correspond to a background given by Eqn.~\ref{eqn:chipg} with $A$ =2.4 10$^{-4}$ emu/mole and $E_G/k_B$ = 413~K. Values of the fitting parameter ``fract'', the fractional length of the arcs affected by the CDW are also shown. }
\label{XRDcomp}
\end{figure}

\begin{figure}
\includegraphics[width=7.5cm,keepaspectratio=true]{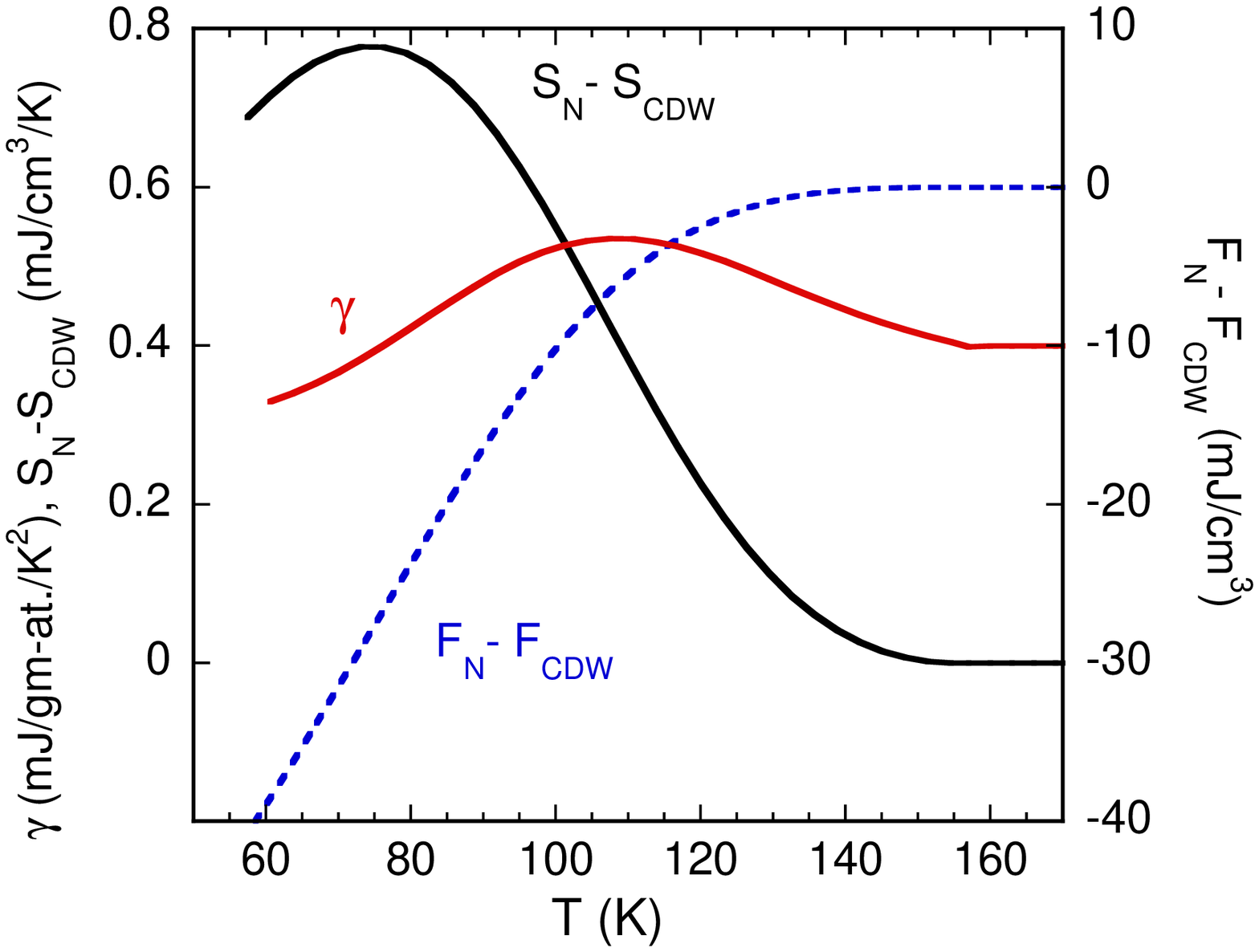}
\caption{Color online.  Calculated changes in electronic entropy $S_N-S_{CDW}$ of UD68 caused by the onset of the CDW below $\simeq$~150~K. The contributions to  $S_N$ and $S_{CDW}$ from the Fermi arcs were calculated using the same parameters that give a good  fit to the data for UD68 in Fig.~\ref{XRDcomp}. The  electronic heat capacity was then found from the thermodynamic relation ($\gamma =C_{el}/T= dS/dT$) and the difference in Helmholtz free energies $F_N-F_{CDW}$ from $F=-dS/dT$. Absolute magnitudes are obtained by normalizing to the experimental value for  YBa$_{2}$Cu$_{3}$O$_{7}$ where there is no pseudogap, $\gamma$ = 2 mJ/gm-at./K$^2$~\cite{Loram01}.}
\label{FreeEnergy}
\end{figure}

The entropy, specific heat and free energy differences calculated in this model, with the same parameters used for the best fit in Fig.~\ref{XRDcomp}, are shown in Fig.~\ref{FreeEnergy}.
We see that at 100~K $F_N - F_{CDW}$ = $10$mJ/cm$^3$ for UD68. Taking the average in-plane coherence length of the CDW, $\xi_{CDW}$ to be 6~nm at 100~K~\cite{Blanco}, together with the $c$-axis spacing $d=1.17$~nm gives the free energy difference multiplied by  the appropriate  coherence volume, $\xi_{CDW}^2d$, as only $30$~K at 100~K and rising to $\sim$70~K at 80~K. This  strongly suggests that thermal fluctuations are highly significant, in agreement with the suggestion in Ref.~\onlinecite{Wu2015}. This estimate does depend on the method used to subtract the background contribution when making the fit to the UD68 data. As shown in Fig.~\ref{XRDcomp}, for a background based on Eqn.~\ref{eqn:chipg}, which  does not give such a good fit, the effect of the CDW is approximately a factor 2 larger.  There is also uncertainty arising from the possibility that $T_{CDW}$ is significantly higher for our twinned UD68 sample than for the detwinned one studied in Ref.~\onlinecite{Chang}.
Together these two uncertainties could increase the effect of the CDW by up to a factor 3, but even then our estimate of $F_N - F_{CDW}$ will not exceed $30$mJ/cm$^3$ so the free energy in a coherence volume at 100~K is definitely less than $k_BT$ at 100~K.
As shown in Fig.~\ref{FreeEnergy} the smallest estimate of the change in free energy below $T_{CDW}$,  gives a broad peak in $\gamma$ near 110~K that is calculated to be 0.14 mJ/gm-at./K$^2$. This is about 8$\%$ of the electronic term for a UD67 sample near 110~K~\cite{Loram01}. We expect there to be a $T$-dependent background caused by the pseudogap, but the effect might possibly be detectable using the ratio method~\cite{Loram01}.
  Such an experiment should be attempted because in principle the heat capacity (and  the thermal expansion) could give larger effects than what we  predict
   from the magnetic susceptibility. For example  (a) from new Raman-active modes attributed to the CDW~\cite{Raman}  or (b) because the onset of the CDW induces a change
    in the pseudogap energy.  Possibility (b) could be important for understanding the driving force for CDW formation. Taking the viewpoint that the pseudogap does not
    conserve states~\cite{Loram01} but instead causes a shift in spectral weight by energies of 1200~K or more away from the chemical potential, then changes in $E_G$ could
   \begin{table}
\begin{ruledtabular}
\begin{tabular}{l|c|c|c|c|c|c}
$wt.$&$\delta m$&$x$&$T_{c}$&$10^4C$&$E_{G}/k_B$&$p$\\
$mg$&$mg$&$meas$&$(K)$&$\frac{emu-K}{mole}$&$(K)$&/CuO$_2$\\
\hline $85.83$&$1.05$&$0.81$ &$84.9$&$33\pm20$&311&0.141\\
\hline$87.04$&$0.96$&$0.76$ &$76.8$&$25\pm10$&353&0.134\\
\hline $85.62$&$0.84$&$0.71$ &$71.9$&$16\pm10$&384&0.129\\
\hline $104.51$&$0.93$&$0.67$ &$67.9$&$42\pm20$&413&0.125\\
\hline $254.13$&$2.01$&$0.63$ &$64.0$&$84\pm20$&476&0.115\\
\hline $131.42$&$1.01$&$0.62$ &$61.9$&$135\pm20$&512&0.109\\
\hline $234.13$&$1.12$&$0.50$ &$57.0$&$50\pm15$&578&0.099\\
\end{tabular}
\end{ruledtabular}
\caption{Summary of data. The first two columns show the final sample weights and the weight gain relative to the state with $x$ = 0.3, giving the final values of $x$ shown.
  The critical temperature $T_c$ was taken from the onset of diamagnetism as determined by measuring  field-warming magnetization at 10 Oe after zero-field cooling. $C$ is the Curie constant and   $E_{G}/k_B$ is the pseudogap energy given by 1200$[1-(p/0.19)]$. The values of $p$ are obtained from the $T_c(p)$ curve for annealed crystals in  Ref.~\onlinecite{Liang}.}
 \label{summary}
\end{table}
    alter the electronic free energy significantly  and therefore affect the heat capacity, without necessarily showing up in the magnetic susceptibility. However
    we have no explanation as to how the CDW could cause changes in $E_G$ since in a simple picture it would only cause a symmetric spatial modulation about an average
    value. Also no heat capacity anomalies arising from CDWs were seen in heat capacity studies of polycrystalline YBCO samples~\cite{Loram01}.
\section{IV. SUMMARY AND CONCLUSIONS}
We have measured the anisotropic magnetic susceptibility of 7 crystals of YBa$_{2}$Cu$_{3}$O$_{6+x}$ and used a special method for separating out the Curie term caused by a small concentration of localized spins that are induced by disorder on the Cu-O chains. We show that this term flattens out above 200~K for all samples and discuss two possible mechanisms for this unusual effect.
 By assuming that our crystals have the same CDW onset temperatures as those measured by resonant X-ray scattering we can detect small effects in the magnetic susceptibility associated with the formation of charge density waves. These are stronger for samples on the $T_c(x)$ plateau, supporting scenarios in which the charge density waves are the main cause of the plateau. The effects are very small, which we ascribe, at least partly, to  the pseudogap reducing the electronic entropy and density of states above $T_{CDW}$. For UD68 we have analyzed them  using a minimal model involving   a cylindrical Fermi surface and short Fermi arcs. This analysis shows that the CDW  causes a reduction in  electronic free energy in  a coherence volume that is less than $k_BT$ at 100~K and comparable with, or up to a factor 2 larger than, $k_BT$ at 80~K. This is consistent with  CDW being  a bulk effect, i.e. not necessarily nucleated around defects~\cite{Wu2015,BlancoZn}, but with a slow onset that is probably caused by thermal fluctuations.  The Wilson ratio for the cuprates is similar to that expected for weakly interacting fermions~\cite{Cooper1996}.  Therefore comparable values for the electronic free energy would be obtained using a more realistic band structure and  different arc lengths, because of the general relation between the  measured magnetic susceptibility and  the electronic heat capacity. Estimates of the size of the CDW gap  suggest that there are unlikely to be significant CDW-induced changes in transport properties at 100~K at low magnetic fields, i.e. the Hall coefficient would be hole - rather than electron-like in agreement with experimental data~\cite{LeBoeufHall,Walker,AndoHall}.  They also imply that, in the first approximation, all transport properties at low magnetic  fields should be discussed in terms of Fermi arcs~\cite{Gorkov}  rather than pockets of carriers.
 We have also put forward a qualitative explanation for the enhancement of the CDW by a large magnetic field. This involves a pile up of charge at the tips of the Fermi arcs that is induced by the Lorentz force. It remains to be seen whether this could be confirmed  by calculating the magnetic field dependence of the $Q$-dependent charge susceptibility $\chi(Q)$ for Fermi arcs. One possible approach might be  along the same lines as the calculations~\cite{Montambaux} of $\chi(Q)$  that account for  field induced spin density waves in the quasi-one-dimensional Bechgaard salts.
\section{ACKNOWLEDGEMENTS}
We are grateful to K. Iida for supplying us with a 1.5 gm YBa$_{2}$Cu$_{3}$O$_{6.3}$ crystal and for information about its preparation and to   J. C. Baglo, A. Carrington, A. J\'{a}nossy, J. W. Loram, J. L. Tallon and E. Tuti\v{s} for helpful discussions.
 This work has been supported  by the Croatian Science Foundation under the project (No. 6216), by the Croatian Research Council, MZOS NEWFELPRO project No. 19 and by  the Engineering and Physical Sciences
Research Council (UK) (grant no. EP/K016709/1).


\begin{thebibliography}{99}
\bibitem{Crystaldetails}    UD85, 7 days in flowing air (A) at 540~C, then 10 months at 25~C (RT). UD77, 2.7 days A  555~C, 30 months RT.  UD72, 7 days A  570~C, 2 months RT.   UD68, 5 days A 590~C, 2 months RT. UD64, 6 days A 610~C, 1 month RT. UD62, 6 days A 640~C 10 months RT. UD57, 8 days  flowing 1$\%$ O$_2$/N$_2$  560 C, 2 months RT.
\end{thebibliography}
\end{document}